\documentclass[conference]{IEEEtran}
    \IEEEoverridecommandlockouts
    \usepackage{cite}
    \usepackage{amsmath,amssymb,amsfonts}
    \usepackage{algorithmic}
    \usepackage{graphicx}
    \usepackage{textcomp}
    \usepackage{xcolor}
    \usepackage{booktabs}
    \usepackage{caption}
    \usepackage{flushend}
    \usepackage{listings}
    \usepackage{xcolor}
    
    \def\BibTeX{{\rm B\kern-.05em{\sc i\kern-.025em b}\kern-.08em
        T\kern-.1667em\lower.7ex\hbox{E}\kern-.125emX}}
    
\begin{document}

    \title{Copper: Unifying Correctness and Performance Specification in Code Generation}


    \author{
    \IEEEauthorblockN{André Lizardo}
    \textit{Red Hat, Madrid, Spain}\\
    \textit{University of Coimbra, CISUC, DEI, Portugal}\\
    alizardo@redhat.com
    \and
    \IEEEauthorblockN{Raul Barbosa}
    \textit{University of Coimbra, CISUC, DEI}\\
    Coimbra, Portugal \\
    rbarbosa@dei.uc.pt
    }
    
    \maketitle

    \begin{abstract}
    Generative AI has made remarkable progress in producing functionally correct code, yet ensuring both correctness and performance remains an open challenge. We present Copper, a framework that combines formal verification with performance-aware specification to generate code that is provably correct and efficiently executable. Our approach integrates AI-driven code synthesis with formal verification tools, and automated performance profiling loops. Evaluated on a diverse set of algorithmic and real-world programming tasks, Copper produces solutions that satisfy strict correctness guarantees while delivering significant improvements in runtime and memory efficiency compared to baseline AI-generated code. This work demonstrates that it is feasible to bridge the gap between trustworthiness and performance in AI-assisted programming, offering a practical pathway toward reliable, high-performance code generation.
    \end{abstract}

    \begin{IEEEkeywords}
    Software Engineering, Generated Code, Formal Verification, Correctness, Performance Analysis
    \end{IEEEkeywords}

    \section{Introduction}



Although code generation has advanced significantly, Large Language Models (LLMs) still lack the ability to generate reliable, verifiable code. Moreover, while correctness has received considerable attention, code performance has been largely overlooked, despite the fact that both are necessary. This paper aims to guide the generation process toward code that is not only correct but also satisfies specified complexity constraints.

Generative AI can produce correct code for a variety of tasks, but guaranteeing both correctness and performance remains a challenge. Traditional optimizations rely on human insight or post-hoc tuning, leaving AI-generated solutions correct but inefficient. This paper introduces Copper (short for ``Correctness Plus Performance'') which is a framework that bridges this gap by integrating formal verification with performance-aware code generation, ensuring correctness and efficiency simultaneously. It uses Dafny~\cite{leino2023program} as an intermediate language, allowing the generated code to be formally verified.

Software development is undergoing a fundamental transformation. Just as the introduction of Fortran in the late 1950s allowed developers to move up from assembly to high-level programming languages, the field is now witnessing a further shift in abstraction, from high-level languages to specifications.

The goal is for LLMs to generate trustworthy code, ideally paired with formal specifications which can be verified in an automated way. In practice, however, LLMs remain unable to generate formally verified code autonomously. This limitation defines the research problem this paper addresses.

In the proposed approach, Dafny serves as the intermediate language, enabling specification-driven code generation. Generative AI is prompted to produce a formal specification, including a proof of algorithmic complexity, from which the corresponding implementation is derived and verified. The Dafny backend can then transpile the result to languages such as Python for final deployment. Correctness is specified through contracts, which are formally verified to hold for the generated implementation. Computational complexity, meanwhile, serves as a formally verifiable proxy for performance, with the complexity proof embedded directly in the specification.

The first contribution is the design of a generation pipeline consisting of four stages: prompt construction and submission to the model, parsing and validation of the response into a Dafny file, formal verification via the Dafny verifier, and estimation of the time complexity of the final Python code. The pipeline is iterative, running up to a specified maximum number of attempts, with verification errors fed back into the loop to guide repair of the generated code.

The second contribution is the design and comparative evaluation of three prompt variants. A short prompt, serving as the baseline, provides a brief natural-language statement of the specification. A detailed prompt supplements this with additional information to clarify the specification. A formal prompt goes further, presenting the specification as a Dafny contract, namely, the requires and ensures clauses that the generated implementation must satisfy.

The third contribution is the design of an agent architecture and composition strategy for guiding code generation toward correct, formally verified solutions, together with an evaluation across a range of language models. The proposed approach improves the success rate from a low and variable pass@1 to a substantially higher repair@10.

The remainder of the paper is organized as follows. Section~\ref{sec:related} contextualizes the proposed approach and describes related work. Section~\ref{sec:method} describes the proposed approach. Section~\ref{sec:results} presents the results of the experimental evaluation. The conclusions are described in Section~\ref{sec:conclusion}.

    \section{Related Work}\label{sec:related}



    Prior research on LLM-generated code has converged into four main paradigms.
    
    \begin{itemize}
        \item \textbf{Empirical Validation} --- assess correctness by developing benchmarks, containing problems and test suites such as HumanEval, SWE-bench or FeedbackEval.
        \item \textbf{Empirical Validation with feedback} --- assesses correctness by running feedback loops where code was proven to be incorrect (using unit tests) and re-submitting it back to the model for repair.
        \item \textbf{Formal Verification} --- assess correctness verifiable artifacts validated against explicit specifications (pre-conditions, post-conditions, and invariants).
        \item \textbf{Performance Benchmarks} --- assess the performance by measuring the execution time and CPU instruction counts.
    \end{itemize}
    
    While benchmarks like HumanEval\ \cite{chen2021evaluating} and SWE-bench have become standard for assessing functional correctness through test suites. Functional validation may be insufficient due to lack of coverage and risk of data contamination - LLMs can game the benchmark \cite{wangtoward, dai2025feedbackeval}. Addressing such limitations, Dai et al.\ \cite{dai2025feedbackeval} introduced FeedbackEval, a framework that evaluates and repairs generated code through feedback loops. Moreover, Li et al.\ \cite{li2025dafny} introduced a Dafny Pipeline. It generates Dafny code as intermediate representation that can be verified against agreed specifications.
    
    Unlike functional validation, formal verification provides an absolute guaranty of the correctness of a given specification. Regarding performance, efficiency is measured through execution time \cite{coignion2024performance} or CPU instruction count \cite{peng2025coffe}. However, such metrics are highly sensitive to the hardware and runtime environment. 

    \begin{table}[htbp]
    \centering
    \caption{Comparison of the different approaches for evaluating correctness and performance of LLM-generated code}
    \label{tab:eval_paradigms}
    \begin{tabular}{lcccc}
        \toprule
        \textbf{Solution} & \textbf{Correctness} & \textbf{Performance} & \textbf{Repair} \\
        \midrule
        HumanEval \cite{chen2021evaluating} & Unit tests & No & No \\
        SWE-bench \cite{jimenez2023swe} & Project tests & No & No \\
        FeedbackEval \cite{dai2025feedbackeval} & Unit tests & No & Yes \\
        Dafny-Pipeline \cite{li2025dafny} & Dafny Verifier & No & No \\
        \midrule
        LLM-Leetcode \cite{coignion2024performance} & Unit tests & Execution & No \\
        & & Runtime & \\
        \midrule
        COFFE \cite{peng2025coffe} & Unit tests & CPU & No \\
        & & Instruction Count & \\
        \midrule
        \textbf{Copper} & \textbf{Dafny Verifier} & \textbf{Time Complexity} & \textbf{Yes} \\
        & & \textbf{Bound Postcondition} & \\
        \bottomrule
    \end{tabular}
    \end{table}

    As demonstrated in Table~\ref{tab:eval_paradigms}, current research focus mainly on empirical validation using unit-testing and on classic performance benchmarks by measuring execution time. 
    
    To address such limitations, we introduce Copper (Complexity Plus Performance) an evaluation pipeline which applies formal verification of correctness and performance to evaluate generated code. 
    \section{Method}\label{sec:method}

    Copper evaluates whether generated code can be formally correct and asymptotically efficient. This pipeline is composed by five discrete stages:

    \begin{itemize}
        \item \textbf{PromptBuilder} --- Prepares, enhances and builds the prompt before requesting the model.
        \item \textbf{ModelRequester} --- Requests the model and parses the model response, validating whether the response contains a source code block and parsing it into a Dafny code file.
        \item \textbf{CorrectnessVerifier} --- Runs Dafny Verify assessing correctness and performance, then runs Dafny Build generating Python code from the LLM-generated code. In formal specifications, CorrectnessVerifier assesses performance verifying the time-complexity postcondition. 
        \item \textbf{PerformanceAnalyzer} --- Estimates the time-complexity of the Python code.
        \item \textbf{SanityChecker} --- Runs the Python code using provided inputs and asserts the results with provided expected results.
    \end{itemize}

    Each evaluation executes the entire pipeline in a fixed order, starting from a prompt preparation up to running a sanity check.
    
    \subsection{Prompt types.} 
    Copper was developed to handle three different prompt scenarios. 
    
    \textbf{Short prompts} represent a short phrase, as shown by Figure \ref{fig:short-prompt}, the typical day-to-day request that a developer makes, containing a small representation of the task and expectations but lacking detailed specifications.
    
    \begin{figure}[!t]
    \small

    \begin{quote}
    \small
    \textit{Solve the following problem in Dafny: given a sorted array
    \texttt{array\_A} and a target \texttt{target\_B}, return the index of \texttt{target\_B}, or -1 if \texttt{target\_B} is absent. \ldots}
    \\
    \textit{Requirements:}
    \begin{itemize}
          \item The solution must be verifiable and have \texttt{O(log N)} time complexity
      \item The solution method must be named "exec", take two arguments "array\_A" (the array) and "target\_B" (the target), and return an integer
    \end{itemize}
    \end{quote}
    
    \captionof{figure}{Short prompt for the search problem.}
    \label{fig:short-prompt}
    \end{figure}
    
    \textbf{Detailed prompt} improve upon the short prompt, as shown by Figure \ref{fig:detailed-prompt}. It contains more requirements, expectations, edge cases, scenarios written in natural language. This represents the prompts used for "vibe-coding" applications.

    \begin{figure}[!t]
    \small
    \begin{quote}
    \small
    \textit{Solve the following problem in Dafny: given a sorted array
    \texttt{array\_A} and a target \texttt{target\_B}, return the index of
    \texttt{target\_B}, or \texttt{-1} if absent.}
    
    \textit{Requirements:}
    \begin{itemize}
      \item The solution must be verifiable and meet \texttt{O(log N)} time complexity
      \item The solution method must be named \texttt{exec}, take two arguments \texttt{array\_A} and \texttt{target\_B}, and return an integer.
      \item Return only Dafny source code---no markdown, preamble, or explanation.
    \end{itemize}
    
    \textit{Dafny refinement:}
    \begin{itemize}
      \item Development: State \texttt{requires} and \texttt{ensures} first, then implement; add loop \texttt{invariant} and \texttt{decreases} as needed
      \item Method: Use exactly the signature \texttt{exec(array\_A: array<int>, target\_B: int) returns (index: int)} ...
      \item Verification: single complete, compilable .dfy program ...
      \item Requires: state all preconditions the solution relies on.  ...
      \item Ensures: fully specify the contract of the return value. ...
      \item (more details were omitted)
    \end{itemize}
    \end{quote}
    
    \captionof{figure}{Detailed prompt for the search problem (partially omitted).}
    \label{fig:detailed-prompt}
    \end{figure}
    
    \textbf{Formal prompts} represent formal specifications where the developer provides the formal contract written in Dafny containing method signature, preconditions and postconditions.

    \begin{figure}[!t]
\small
\begin{quote}
\small
\textit{Complete the following method body:}

\begin{verbatim}
ghost function Log2(n: nat): nat
  requires n > 0
  decreases n
{
  if n <= 1 then 0 else 1 + Log2(n / 2)
}

ghost var steps: int

method exec(arr_A: array<int>, tg_b: int)
  returns (index: int)
  reads arr_A
  requires forall i, j :: ...
           ==> arr_A[i] <= arr_A[j]
  ensures index == -1 ==> forall i :: ...
           arr_A[i] != tg_b
  ensures index >= 0 ==> 0 <= ...
           && arr_A[index] == tg_b
  ensures steps == if arr_A.Length <= 1
           then arr_A.Length
           else Log2(arr_A.Length) + 1
{
}
\end{verbatim}
\end{quote}

\captionof{figure}{Formal prompt for the search problem (partially omitted).}
\label{fig:formal-prompt}
\end{figure}
    
    \subsection{Single-attempt execution.} 
    \textit{1. PromptBuilder} ingests a classic textbook problem-specific prompt with the expected time complexity, for instance: \textit{Solve the following problem in Dafny: given array\_A of integers, return the same elements rearranged in non-decreasing order by meeting the time complexity O(n)}. If repair mode is enabled, it also enhances the prompt with previous generated code and errors. \textit{2. ModelRequester} requests the model using such prompt, extracts the Dafny code block from the payload and persists it as a \textit{.dfy} file. \textit{3. CorrectnessVerifier} runs \emph{dafny verify} on that file to formally check correctness and performance, then runs \emph{dafny build} to generate \textit{Python} code. \textit{4. PerformanceAnalyser} executes the \textit{Python} code and estimates its asymptotic time-complexity comparing it with the prompt expectation. Finally, \textit{5. SanityChecker} runs the python code with a fixed input and tests if the results match the expected output.
    
    \subsection{Feedback-Repair loop.}
    When repair mode is enabled, attempts are no longer independent, they share contexts. Each step failure is reported back to the attempt execution history. The \textit{PromptBuilder} injects the full attempt history by appending the previous attempts context (generated code, exceptions, failures and successes) to the prompt. The prompt is sent to the model containing context of the previous runs, helping the model understanding why the generated code failed to pass a specific step of the pipeline. The repair loop finishes when all pipeline stages succeed or the maximum number of attempts is reached.
    
    \subsection{Correctness Approach.} 
    This approach is a key design choice in the Copper architecture. Guides the model toward better results, improving both correctness and performance.
    
    \textit{Dafny} formally verifies the generated code. Such verification includes checks for preconditions, postconditions, loop invariants and terminations. For formal prompts, the time-complexity is also bound to contract as a postcondition (ghost step counter) which forces the model to generate code where efficiency can be formally assessed. For instance, to force a linear time complexity, the postcondition can be set as: \emph{ensures steps == array\_A.Length} followed by a ghost variable \emph{ghost var steps: int}. This is further detailed in Section~\ref{sec:ghost}.

    By unifying correctness and performance at the level of formal proof, a new research dimension is introduced: models must satisfy correctness and performance without sacrificing any of them. Beyond that, introducing such design into the formal specification allows optimisation of performance and correctness -- much like GCC optimisation flags -- allowing researchers to relax or tighten verification requirements trading efficiency against correctness or vice versa.
    
    \subsection{Performance checks.}
    As mentioned above, formal prompts can enforce performance through formal verification. Apart from that, time-complexity is estimated and validated against benchmarked results. \textit{PerformanceAnalyzer} runs and estimates the Python code (built by Dafny) validating their estimation results. In short, performance check contains two main evaluations: formal verification and empirical time estimation.
    
    \subsection{Metrics and Measurements}
    Copper outputs two kind of metrics \textit{pass@k} for measuring if \textit{k} independent shots can produce correct and performant solution, and \textit{repair@k} for measuring if the model can fix bugs in \textit{k} tries. Moreover, \textit{pass@1} measures raw capability, \textit{pass@10} measures consistency, \textit{repair@1} and \textit{repair@10} measure debugging ability. Metrics are averaged over ten repeated iterations (not attempts) to have better statistical stability.

\subsection{Complexity Specification}\label{sec:ghost}

To unify functional correctness and algorithmic complexity in automated code generation, we use Dafny’s ghost state mechanism to instrument code with zero-cost runtime trackers. The provides the means to support the formal equivalent to instruction counting, used in related work~\cite{peng2025coffe} via unit testing as a proxy to performance.

Because deductive verifiers lack a native concept of asymptotic time complexity (such as Big-$\mathcal{O}$ notation), traditional specifications only guarantee what a program computes, leaving how long it takes unverified. Our approach addresses this gap by requiring the code generator to co-synthesize a computational ``budget'' alongside the core logic. This is achieved by introducing explicit ghost step-counter variables that increment at the dominant, high-cost operations of the algorithm, such as the condition checks of a loop or recursive branch invocations.

The core challenge in automated code generation is not merely injecting these counters, but generating the inductive mathematical bridges required to satisfy the SMT solver. For iterative blocks, the generator must emit a ``lock-step'' loop invariant that strictly equates the ghost counter to the progress of the loop indices (e.g., \texttt{invariant steps == current\_index}). Without this explicit relational tie, the solver views the loop as a black box and fails to propagate the tracking data past the loop boundary. By forcing the generator to bind the ghost state to the loop's natural structural boundaries, we transform a vague performance goal into a checkable inductive proof. As show in Figure~\ref{fig:formal-prompt}, the description includes ghost helper functions, such as Log2, to specify the intended complexity, similar to Big-$\mathcal{O}$ notation, which allows one to write the following specification in our running example:

\noindent\texttt{ensures steps == if arr\_A.Length <= 1\\
           then arr\_A.Length\\
           else Log2(arr\_A.Length) + 1}

Furthermore, given that ghost variables are purely architectural artifacts used for verification, they are completely erased during compilation. This provides a clean separation of concerns: the generated software carries a proven guarantee of its resource boundaries during the synthesis phase, yet compiles down into optimized machine code with zero runtime performance penalties.

    \subsection{Methodology}
    To run Copper from the command line, a few parameters are required: prompt, expected algorithm, and model. To enable repair, repair and attempts must also be set.
    
    The prompt is usually passed as a plain text file. The expected algorithm names the textbook algorithm the model should implement and determines the expected time complexity—for example, \textit{binary\_search} has logarithmic complexity. The model parameter sets the model name; when using Ollama, this should be an Ollama model identifier such as \textit{gemma4:31b}. If the experiment uses guided repair loops, repair and attempts must be configured as well. Copper outputs the repair history, raw and base prompts, full model responses, extracted Dafny code, Dafny compiled Python code and a structured event timeline.
    
    Experiments were conducted using the three prompt types (short, detailed and formal) designed to solve a specific algorithmic search problem, where binary search was the expected correct and performant solution. In addition, the experiments were made against three models of different sizes - Gemini 3 Flash Preview (large), Gemma 4 31B (medium), and LLaMA 3 8B (small). Moreover, while large models were hosted in Ollama Cloud services, medium and small models were serve locally by Ollama running on a Kubernetes cluster. Each iteration was represented by a Job resource which spun up a pod with two containers copper and ollama. Finally, the results were persisted to an attached NFS server.

    \section{Results}\label{sec:results}
    The following results represent an experiment designed to solve an algorithmic search problem using binary search. The experiment was run by Copper and went through all pipeline steps. A success output indicates that the solution passed all stages and met the correctness and performance requirements.
    
    Table \ref{tab:full-pipeline-basic-prompt-transposed} shows that \emph{gemini-3-flash-preview} excels: with the short, less restrictive prompt, it solves the problem easily \textit{(pass@1 = 1.0, pass@10 = 1.0, repair@1 = 1.0, repair@10 = 1.0)}. The mid-sized model, \emph{gemma4:31b}, can also produce solutions that meets correctness and performance criteria \textit{(pass@1 = 0.3, pass@10 = 0.9, repair@1 = 0.8, repair@10 = 0.9)}. The smallest model fails drastically, it never generates a successful solution.

    \begin{table}[htbp]
    \centering
    \caption{Evaluation of correctness and performance for binary search solution using short prompt}
    \label{tab:full-pipeline-basic-prompt-transposed}
    \begin{tabular}{lccc}
    \toprule
    Metric & gemini-3-flash-preview & gemma4:31b & llama3-8b \\
    \midrule
    \texttt{pass@1}   & 1.00 & 0.30 & 0.00 \\
    \texttt{pass@10}  & 1.00 & 0.90 & 0.00 \\
    \texttt{repair@1}  & 1.00 & 0.80 & 0.00 \\
    \texttt{repair@10} & 1.00 & 0.90 & 0.00 \\
    \bottomrule
    \end{tabular}
    \end{table}
    
    With a more detailed prompt, the models tend to underperform. Table \ref{tab:full-pipeline-refined-prompt-transposed} shows that \emph{gemini-3-flash-preview} could only produce a successful solution when repair mode was enabled \textit{(repair@1 = 1.0, repair@10 = 1.0)}. The mid-sized model, \emph{gemma4:31b}, had the same behaviour but presenting worse results than the larger model \textit{(repair@1 = 0.0, repair@10 = 0.5)}. The smallest model keeps failing drastically, a more detailed prompt did not improve its performance.
    
    \begin{table}[htbp]
    \centering
    \caption{Evaluation of correctness and performance for  binary search solution using detailed prompt}
    \label{tab:full-pipeline-refined-prompt-transposed}
    \begin{tabular}{lccc}
    \toprule
    Metric & gemini-3-flash-preview & gemma4:31b & llama3-8b \\
    \midrule
    \texttt{pass@1}   & 0.00 & 0.00 & 0.00 \\
    \texttt{pass@10}  & 0.00 & 0.00 & 0.00 \\
    \texttt{repair@1}  & 1.00 & 0.00 & 0.00 \\
    \texttt{repair@10} & 1.00 & 0.50 & 0.00 \\
    \bottomrule
    \end{tabular}
    \end{table}

    With formal prompt, the models struggle and require multiple repair rounds. Table \ref{tab:full-pipeline-refined-prompt-transposed} shows that \emph{gemini-3-flash-preview} produces a successful only after extended repair \textit{(repair@1 = 0.0, repair@10 = 0.9)}. \emph{gemma4:31b} presents stronger single repair iteration \emph{(repair@1 = 0.3)} and full success during extended repair \textit{(repair@10 = 1.0)}. Finally, the smallest model fails completely, achieving no successful solution.
    
    \begin{table}[htbp]
    \centering
    \caption{Evaluation of correctness and performance for binary search solution using formal prompt}
    \label{tab:full-pipeline-pre-pos-conditions-prompt-transposed}
    \begin{tabular}{lccc}
    \toprule
    Metric & gemini-3-flash-preview & gemma4:31b & llama3-8b \\
    \midrule
    \texttt{pass@1}   & 0.00 & 0.00 & 0.00 \\
    \texttt{pass@10}  & 0.00 & 0.00 & 0.00 \\
    \texttt{repair@1}  & 0.00 & 0.30 & 0.00 \\
    \texttt{repair@10} & 0.90 & 1.00 & 0.00 \\
    \bottomrule
    \end{tabular}
    \end{table}

    Copper uses formal verification techniques alongside with empirical time-complexity evaluations to ensure correctness and efficiency simultaneously, as shown in Table~\ref{tab:full-pipeline-pre-pos-conditions-prompt-transposed}. 

    While the main criteria are about guaranteeing both correctness and performance, reporting correctness and performance separately reveal whether the factors (model, prompt) struggle with verification, efficiency, or both.

    \subsection{Evaluating Only Correctness}
    Table~\ref{tab:correctness-mandatory-short} shows that \textit{gemini-3-flash-preview} outperforms the rest of the models in both for independent and guided-repair experiments. Additionally, \textit{gemma4-31b} has an high ability to generate correct code, improving its results for higher \textit{k}. Also, \textit{gemma4-31b} achieves better results when the guided repair is enabled. \textit{Llama3-8b} fails to generate any correct code.
    
    \begin{table}[htbp]
    \centering
    \caption{Evaluation of correctness for binary search solution using short prompt}
    \label{tab:correctness-mandatory-short}
    \begin{tabular}{lccc}
    \toprule
    Metric & gemini-3-flash-preview & gemma4:31b & llama3-8b \\
    \midrule
    \texttt{pass@1} & 1.00 & 0.30 & 0.00 \\
    \texttt{pass@10} & 1.00 & 0.90 & 0.00 \\
    \texttt{repair@1} & 1.00 & 0.80 & 0.00 \\
    \texttt{repair@10} & 1.00 & 0.90 & 0.00 \\
    \bottomrule
    \end{tabular}
    \end{table}

    Detailed prompts include more problem details, requirements, and expectations. Table~\ref{tab:correctness-mandatory-detailed} shows that
    \textit{gemini-3-flash-preview} and \textit{gemma4:31b} rely on guided repair to generate correct code. Moreover, \textit{gemma4:31b} needs at least ten repair rounds to produce correct solutions, and remains well behind \textit{gemini-3-flash-preview}. \textit{Llama3-8b} fails to generate any correct code.
    
    \begin{table}[htbp]
    \centering
    \caption{Evaluation of correctness for binary search solution using detailed prompt}
    \label{tab:correctness-mandatory-detailed}
    \begin{tabular}{lccc}
    \toprule
    Metric & gemini-3-flash-preview & gemma4:31b & llama3-8b \\
    \midrule
    \texttt{pass@1} & 0.00 & 0.00 & 0.00 \\
    \texttt{pass@10} & 0.00 & 0.00 & 0.00 \\
    \texttt{repair@1} & 1.00 & 0.00 & 0.00 \\
    \texttt{repair@10} & 1.00 & 0.50 & 0.00 \\
    \bottomrule
    \end{tabular}
    \end{table}

    Table~\ref{tab:correctness-mandatory-formal} details that \textit{gemini-3-flash-preview} struggles to produce correct code using formal specifications without guided repair, getting a positive result only for \textit{repair@10=0.90}. \textit{gemma4:31b} results are different, it struggles without guided repair but it presents better results than any other model - \textit{repair@1=0.30 and repair@10=0.90}. \textit{Llama3-8b} fails to generate any correct code.

    \begin{table}[htbp]
    \centering
    \caption{Evaluation of correctness for binary search solution using formal prompt}
    \label{tab:correctness-mandatory-formal}
    \begin{tabular}{lccc}
    \toprule
    Metric & gemini-3-flash-preview & gemma4:31b & llama3-8b \\
    \midrule
    \texttt{pass@1} & 0.00 & 0.00 & 0.00 \\
    \texttt{pass@10} & 0.00 & 0.00 & 0.00 \\
    \texttt{repair@1} & 0.00 & 0.30 & 0.00 \\
    \texttt{repair@10} & 0.90 & 1.00 & 0.00 \\
    \bottomrule
    \end{tabular}
    \end{table}
    
    \subsection{Evaluating Only Performance}
    Performance is estimated by running the Python code (built by Dafny) validating the results with expected time-complexities. Table~\ref{tab:performance-mandatory-short} shows that larger models tend to excel, while smaller models fail. \textit{gemini-3-flash-preview} and \textit{gemma4:31b} are able to generate efficient code with short prompts. \textit{Llama3-8b} still fails to generate efficient code.
    
    \begin{table}[htbp]
    \centering
    \caption{Evaluation of performance for binary search solution using short prompt}
    \label{tab:performance-mandatory-short}
    \begin{tabular}{lccc}
    \toprule
    Metric & gemini-3-flash-preview & gemma4:31b & llama3-8b \\
    \midrule
    \texttt{pass@1} & 1.00 & 0.80 & 0.00 \\
    \texttt{pass@10} & 1.00 & 1.00 & 0.00 \\
    \texttt{repair@1} & 1.00 & 1.00 & 0.00 \\
    \texttt{repair@10} & 1.00 & 1.00 & 0.00 \\
    \bottomrule
    \end{tabular}
    \end{table}

    Table~\ref{tab:performance-mandatory-detailed} shows that \textit{gemini-3-flash-preview} requires guided repair to generate efficient code with detailed prompts. This also applies to \textit{gemma4:31b}, although \textit{gemini-3-flash-preview} generates efficient code with a single guided repair iteration (\textit{repair@1 = 1.00}). \textit{Llama3-8b} still fails to generate efficient code.

    \begin{table}[htbp]
    \centering
    \caption{Evaluation of performance for binary search solution using detailed prompt}
    \label{tab:performance-mandatory-detailed}
    \begin{tabular}{lccc}
    \toprule
    Metric & gemini-3-flash-preview & gemma4:31b & llama3-8b \\
    \midrule
    \texttt{pass@1} & 0.00 & 0.00 & 0.00 \\
    \texttt{pass@10} & 0.00 & 0.00 & 0.00 \\
    \texttt{repair@1} & 1.00 & 0.00 & 0.00 \\
    \texttt{repair@10} & 1.00 & 0.60 & 0.00 \\
    \bottomrule
    \end{tabular}
    \end{table}

    With formal prompts, larger models require guided repair iterations to produce efficient code - \textit{gemini-3-flash-preview} requires \textit{k=10} iterations and \textit{gemma4:31b} requires \textit{k=1} improving its results when \textit{k=10}. For formal prompts, \textit{gemma4:31b} outperforms the rest of the models.

    \begin{table}[htbp]
    \centering
    \caption{Evaluation of performance for binary search solution using formal prompt}
    \label{tab:performance-mandatory-formal}
    \begin{tabular}{lccc}
    \toprule
    Metric & gemini-3-flash-preview & gemma4:31b & llama3-8b \\
    \midrule
    \texttt{pass@1} & 0.00 & 0.00 & 0.00 \\
    \texttt{pass@10} & 0.00 & 0.00 & 0.00 \\
    \texttt{repair@1} & 0.00 & 0.70 & 0.00 \\
    \texttt{repair@10} & 0.90 & 1.00 & 0.00 \\
    \bottomrule
    \end{tabular}
    \end{table}

    \subsection{Statistical Analysis: Which prompt--model pairs are statistically better fit for pass@10?}
    
    The results are binomial outcomes, presenting pass or fail per iteration and attempt. Therefore, the main statistical method used is \textit{Chi-Square Test} - it evaluates if there is statistical evidence between categorical variables (pass/fail). 

    \textbf{} To answer this question, it is required to understand if there are pairs that differ statistically in their success rate. For that, the following hypotheses were defined: 
    
    \begin{itemize}
        \item H$_0$: all model--prompt pairs have equal \textit{pass@10} success rates;
        \item H$_1$: there are model--prompt pairs with lower or higher \textit{pass@10} success rate.
    \end{itemize}

    \begin{table}[ht]
      \centering
      \caption{Chi-square tests for \textit{pass@10}.}
      \label{tab:pass10-chisq}
      \begin{tabular}{lrrrl}
        \toprule
        Effect & df & $\chi^2$ & $pvalue$ & \\
        \midrule
        Model         & 2 & 12.1 & $2.31 \times 10^{-3}$  & **   \\
        Prompt        & 2 & 48.2 & $3.47 \times 10^{-11}$ & **** \\
        Model:Prompt  & 8 & 84.6 & $5.78 \times 10^{-15}$ & **** \\
        \bottomrule
      \end{tabular}
    \end{table}
    
    As shown in Table \ref{tab:pass10-chisq}, the omnibus \textit{Chi-Square Test} rejects H$_0$ ($\chi^2 = 84.6 $ and $pvalue = 5.78 \times 10^{-15}$) across all model-prompt pairs. It also reports that there is a strong prompt effect ($\chi^2 = 48.2 $ and $pvalue = 3.47 \times 10^{-11}$) and slight model effect too ($\chi^2 = 12.1 $ and $pvalue = 2.31 \times 10^{-3}$). $\chi^2$ might be inaccurate due to only running ten iterations per experiment.

    
    In order to understand which pairs outperform the others, it is required to do a follow-up analysis. \textit{2×2 chi-square tests} tests the association between two categorical variables.

    \begin{table}[ht]
      \centering
      \caption{Pairwise chi-square tests for \textit{pass@10}.}
      \label{tab:pass10-p-vs-rest}
      \begin{tabular}{lc}
        \toprule
        Model--prompt pair & pvalue\_vs\_rest \\
        \midrule
        gemini + short      & $1.26 \times 10^{-9}$  \\
        gemma4 + short      & $1.51 \times 10^{-7}$  \\
        gemini + detailed   & 0.185                  \\
        gemini + formal     & 0.185                  \\
        gemma4 + detailed   & 0.185                  \\
        gemma4 + formal     & 0.185                  \\
        llama3-8b + detailed & 0.185                 \\
        llama3-8b + formal  & 0.185                  \\
        llama3-8b + short   & 0.185                  \\
        \bottomrule
      \end{tabular}
    \end{table}
    
    Table \ref{tab:pass10-p-vs-rest} details that \textit{gemini + short} ($pvalue = 1.26 \times 10^{-9}$) and \textit{gemma4 + short} with ($1.51 \times 10^{-7}$) present $pvalue < 0.05$ meaning that they have statistically better results for $pass@10$.
    
    \subsection{Statistical Analysis: Which prompt--model pairs are statistically better fit for repair@10?} 
    
    This question requires the same set of tests. The following hypotheses were defined as:

     \begin{itemize}
        \item H$_0$: all model--prompt pairs have equal \textit{repair@10} success rates;
        \item H$_1$: there are model--prompt pairs with lower or higher \textit{repair@10} success rate.
    \end{itemize}

    \begin{table}[ht]
      \centering
      \caption{Chi-square tests for \textit{repair@10}.}
      \label{tab:chisq-repair10}
      \begin{tabular}{lrrrl}
        \toprule
        Effect & df & $\chi^2$ & $pvalue$ & \\
        \midrule
        Model         & 2 & 66.2 & $4.26 \times 10^{-15}$ & **** \\
        Prompt        & 2 &  1.47 & 0.48                    & ---   \\
        Model:Prompt  & 8 & 72.2 & $1.76 \times 10^{-12}$  & **** \\
        \bottomrule
      \end{tabular}
    \end{table}
    
    As shown in Table \ref{tab:chisq-repair10}, \textit{Chi-Square Test} rejects H$_0$  ($\chi^2 = 72.2 $ and $pvalue = 1.76 \times 10^{-12}$) across all model-prompt pairs. Additionally, it reports strong model effect ($\chi^2 = 66.2 $ and $pvalue = 4.26 \times 10^{-15}$). $\chi^2$ might be inaccurate due low amount of iterations per experiment. In short, there are statistical evidence that \textit{model:prompt} have different success rates for \textit{repair@10}, as well as that different models have impact in such metric too.

    \begin{table}[ht]
      \centering
      \caption{Pairwise chi-square tests for \textit{repair@10}.}
      \label{tab:repair10-p-vs-rest}
      \begin{tabular}{llr}
        \toprule
        Model--prompt pair & Success & $p_{\text{vs rest}}$ \\
        \midrule
        gemini + detailed    & 10/10 & 0.0138 \\
        gemini + short       & 10/10 & 0.0138 \\
        gemma4 + formal      & 10/10 & 0.0138 \\
        gemini + formal      & 9/10  & 0.0751 \\
        gemma4 + short       & 9/10  & 0.0751 \\
        gemma4 + detailed    & 5/10  & 0.791  \\
        llama3-8b + detailed & 0/10  & $2.39 \times 10^{-4}$ \\
        llama3-8b + formal   & 0/10  & $2.39 \times 10^{-4}$ \\
        llama3-8b + short    & 0/10  & $2.39 \times 10^{-4}$ \\
        \bottomrule
      \end{tabular}
    \end{table}

    As mentioned in the first question, this test requires a follow-up using \textit{2×2 chi-square tests}, Table \ref{tab:repair10-p-vs-rest} details that \textit{gemini + short prompt}, \textit{gemini + detailed prompt} and \textit{gemma4 + formal prompt} are statistically better having \textit{success = 10/10} and $pvalue = 0.0138$ vs the rest. Therefore, these pair combinations present statistically better results for \textit{repair@10}.

    \section{Discussion}\label{sec:discussion}

    \textbf{Do models perform better with short prompts?}
    Table \ref{tab:pass10-chisq} shows that Prompt has a statistically significant effect on \textit{pass@10} results with short prompts performing best.
    
    This pattern does not hold for \textit{repair@10}: (Table \ref{tab:chisq-repair10}), Prompt does not have a statistically significant prompt effect. Table \ref{tab:pass10-p-vs-rest} further indicates that larger models perform statistically better with short prompts only in Independent runs.

    In short, results suggest that larger models prefer prompts with less restrictive prompts---but only under independent runs.

    \textbf{Do models perform better when correctness is ignored but performance is not?}
    Yes. Larger models tend to achieve higher \texttt{repair@k} when correctness is not required. Table \ref{tab:performance-mandatory-formal} shows that formal prompts have higher \textit{repair@k} scores when ignoring correctness (Table \ref{tab:full-pipeline-pre-pos-conditions-prompt-transposed}) mainly \textit{gemma4:31b} at \texttt{repair@1}.

    \textbf{What is the impact of increasing $k$?}
    $k$ denotes the number of attempts per iteration. Its effect is less noticeable in independent runs, like \textit{pass@k}. In guided repair,a larger $k$ allows more repair loops leading to increased model reasoning which results to higher success rates.

    \textbf{Why models underperform with formal prompts?}
    As shown in Tables \ref{tab:correctness-mandatory-formal}, \ref{tab:full-pipeline-pre-pos-conditions-prompt-transposed}, formal prompts require the model to follow a predefined formal specification. The model must understand the preconditions and postconditions before starting writting the code, which demands additional reasoning rather than a simple pattern-matching exercise. Moreover, time-complexity is also bound as a postconditions further increasing task difficulty. Ultimately, even when the model generates a syntactically valid Dafny code, it must still pass full verification.
    
    \textbf{Do guided repair attempts improve the correctness and performance of generated code when compared to independent attempts?} 

    Initial results show that large models (\textit{gemini3-flash-preview, gemma4-31b}) tend to struggle with detailed and formal prompts. More specifically, they tend to struggle more generating correct code (Tables \ref{tab:correctness-mandatory-short}, \ref{tab:correctness-mandatory-detailed} and \ref{tab:correctness-mandatory-formal}) than efficient code (Tables \ref{tab:performance-mandatory-short}, \ref{tab:performance-mandatory-detailed} and \ref{tab:performance-mandatory-formal}). Moreover, smaller models fail entirely regardless the prompt or of whether independent or guided-repair runs are used.

    In addition, Tables \ref{tab:full-pipeline-basic-prompt-transposed}, \ref{tab:full-pipeline-refined-prompt-transposed} and \ref{tab:full-pipeline-refined-prompt-transposed} show larger models achieve better results using guided repair iterations than independent runs. This is an expected behavior because the guided repair feedbacks attempts logs, errors and generated code to the prompt giving more context for model reasoning, improving results. 

    Table \ref{tab:repair10-p-vs-rest} shows that the top-tier model-prompt pairs include larger models only for \textit{repair@10}. Comparing it with \textit{pass@10} using Table \ref{tab:pass10-p-vs-rest}, it can be concluded that larger models, regardless the prompt type used, have better results when use guided repair instead of independent runs.
    
    \section{Conclusion}\label{sec:conclusion}

This paper addressed the challenge of guiding LLMs toward generating code that is both functionally correct and performant. To this end, Copper was proposed, combining formal verification with complexity-aware code generation using Dafny as an intermediate language. Three prompt strategies were designed and evaluated, and an architecture was developed to guide the generation process iteratively toward correct, formally verified solutions. Experiments show that the proposed approach substantially improves success rates over a single-pass baseline.

The experimental results confirm that a single prompt is insufficient to obtain a correct, formally verified solution, given that pass@1 is effectively zero across nearly all executions. In a substantial number of cases, the LLM asserts that the generated implementation is formally verified, while the Dafny verifier rejects it outright, supporting the need for a guided, iterative approach.

Model size proves to be a determining factor: models up to 8 billion parameters consistently fail to produce valid solutions, while larger models achieve a reasonable success rate. Nevertheless, most models exhibit notable limitations in generating Dafny code, likely due to its niche status and limited representation in training data.

Repair mode consistently improves results across all models, substantially increasing the success rate for generating code that satisfies both functional and complexity specifications, regardless of the prompt used. This confirms that feeding verification errors back into the generation loop is an effective strategy for guiding models toward correct, performant solutions.

Among the three prompt variants, the short prompt yields the best results, followed by the formal prompt, with the detailed prompt performing worst. This suggests that current LLMs perform substantial internal processing of the prompt prior to generating a response, such that an extensively detailed prompt counterintuitively reduces performance. Similarly, although the formal prompt constrains the solution space through a precise mathematical formulation, models fail to leverage this structure to their advantage.

The proposed approach demonstrates that combining formal verification with iterative, feedback-driven generation is a viable path toward trustworthy code synthesis. Copper provides a practical framework that can be applied across a range of tasks and language models, offering a principled mechanism for enforcing both functional correctness and complexity guarantees without relying solely on model confidence. The results also point to clear directions for future work. The limited availability of Dafny in model training corpora remains a significant obstacle, and progress in this area will likely depend on the curation of dedicated training datasets and the fine-tuning of models on high-quality Dafny examples. As formal verification languages gain traction in AI-assisted development, such efforts may prove essential to realizing the full potential of specification-driven code generation.











\section*{Acknowledgement}
This work is funded by national funds through FCT -- Foundation for Science and Technology, I.P., through the Digital Person project (Ref.\ 14765, COMPETE2030-
FEDER-00926800) and within the scope of the research unit UID/00326 -- Centre for Informatics and Systems of the University of Coimbra.

    \bibliographystyle{IEEEtran}
    \nocite{*} 
    \bibliography{references}

@article{he2025swe,
  title={Swe-perf: Can language models optimize code performance on real-world repositories?},
  author={He, Xinyi and Liu, Qian and Du, Mingzhe and Yan, Lin and Fan, Zhijie and Huang, Yiming and Yuan, Zejian and Ma, Zejun},
  journal={arXiv preprint arXiv:2507.12415},
  year={2025}
}

@article{leong2024,
  title={Translating meaning representations to behavioural interface specifications},
  author={Leong, Iat Tou and Barbosa, Raul},
  journal={Journal of Systems and Software},
  volume={211},
  pages={112009},
  year={2024},
  publisher={Elsevier}
}

@article{jimenez2023swe,
  title={Swe-bench: Can language models resolve real-world github issues?},
  author={Jimenez, Carlos E and Yang, John and Wettig, Alexander and Yao, Shunyu and Pei, Kexin and Press, Ofir and Narasimhan, Karthik},
  journal={arXiv preprint arXiv:2310.06770},
  year={2023}
}

@article{chen2021evaluating,
  title={Evaluating large language models trained on code},
  author={Chen, Mark and Tworek, Jerry and Jun, Heewoo and Yuan, Qiming and Pinto, Henrique Ponde De Oliveira and Kaplan, Jared and Edwards, Harri and Burda, Yuri and Joseph, Nicholas and Brockman, Greg and others},
  journal={arXiv preprint arXiv:2107.03374},
  year={2021}
}

@article{dai2025feedbackeval,
  title={Feedbackeval: A benchmark for evaluating large language models in feedback-driven code repair tasks},
  author={Dai, Dekun and Liu, MingWei and Li, Anji and Cao, Jialun and Wang, Yanlin and Wang, Chong and Peng, Xin and Zheng, Zibin},
  journal={arXiv preprint arXiv:2504.06939},
  year={2025}
}

@article{li2025dafny,
  title={Dafny as verification-aware intermediate language for code generation},
  author={Li, Yue Chen and Zetzsche, Stefan and Somayyajula, Siva},
  journal={arXiv preprint arXiv:2501.06283},
  year={2025}
}

@inproceedings{coignion2024performance,
  title={A performance study of llm-generated code on leetcode},
  author={Coignion, Tristan and Quinton, Cl{\'e}ment and Rouvoy, Romain},
  booktitle={Proceedings of the 28th international conference on evaluation and assessment in software engineering},
  pages={79--89},
  year={2024}
}

@article{peng2025coffe,
  title={Coffe: A code efficiency benchmark for code generation},
  author={Peng, Yun and Wan, Jun and Li, Yichen and Ren, Xiaoxue},
  journal={Proceedings of the ACM on Software Engineering},
  volume={2},
  number={FSE},
  pages={242--265},
  year={2025},
  publisher={ACM New York, NY, USA}
}

@article{wangtoward,
  title={Toward Automated, Contamination-free Dafny Benchmark Generation},
  author={Wang, Changjie and Scazzariello, Mariano and Kosti{\'c}, Dejan and Chiesa, Marco}
}

@book{leino2023program,
  title={Program proofs},
  author={Leino, K Rustan M},
  year={2023},
  publisher={MIT Press}
}

    \end{document}